\documentclass[11pt]{iopart}

\expandafter\let\csname equation*\endcsname\relax
\expandafter\let\csname endequation*\endcsname\relax
\usepackage{epstopdf}
\usepackage{epsf}
\usepackage{epsfig}
\usepackage{cite}
\usepackage{amsmath}
\usepackage{graphicx}
\usepackage{graphics}
\usepackage{iopams}
\usepackage{amsfonts}
\usepackage{setspace}
\usepackage{caption}
\usepackage{multirow}
\newcommand{\tev}{\, \text{TeV}}
\newcommand{\gev}{\, \text{GeV}}

\newcommand{\la}{\langle}
\newcommand{\ra}{\rangle}

\usepackage{lineno}
    
\begin{document}
\title{ $W' \to hH^{\pm}$ decay  in $G(221)$ models}
\author{A Jinaru$^{1,2}$, C Alexa$^1$, I Caprini$^1$ and A Tudorache$^1$}
\address{$^1$ \it{Horia Hulubei National Institute of Nuclear Physics and 
Engineering, P.O. Box MG-6, RO - 077125, Bucharest - M\u agurele, Romania,} }
\address{$^2$ \it{University of Bucharest, Faculty of Physics, 405, Atomi\c stilor Street, RO-077125, M\u agurele, Romania} }
\ead{\mailto{adam.jinaru@nipne.ro}, \mailto{calin.alexa@nipne.ro}, \mailto{caprini@theory.nipne.ro}, \mailto{atudorache@nipne.ro}}
\begin{abstract}
Based on extended  $SU(2)_1 \times SU(2)_2 \times U(1)_\text{X}$ gauge groups, $G(221)$ models predict the existence of an additional heavy gauge boson $W'$ and of a heavy charged Higgs boson $H^\pm$. Within this model we calculate the coupling of $W'$ to  the $hH^{\pm}$ pair, where $h$ can be identified with the Standard Model Higgs boson discovered recently at LHC. Using a  phenomenological constraint on the ratio of the symmetry breaking scales in $G(221)$ models, the $W' \to hH^{\pm}$ decay width and the $hH^{\pm}$ production cross section via an intermediate $W'$ in $pp$ collisions at $\sqrt{s} =$ 8 and 14 TeV are calculated. 
Fiducial cross sections obtained with $G(221)$ models for several final states produced at LHC through the $W' \to hH^{\pm}$ decay are compared with recent results in searches for supersymmetry published by the ATLAS Collaboration.\end{abstract}
\maketitle

\newpage
\section{Introduction}\label{sec:intro}
The  discovery at LHC \cite{higgs} by both ATLAS and CMS  of a new massive scalar particle $h$ with properties closely resembling those of the Higgs boson of Standard Model (SM) stimulated the investigation of more general models which predict several, neutral and charged, Higgs bosons, and which should accomodate in a consistent way this  discovery.

The $G(221)$ models \cite{naturalLRSM,isoconjugateLRSM,prD23,baryon,phobic1,phobic2,phobic3,senjanovic,nu2,fogli1985,kayser,uu1,fogli1991,nu1,nu3,mohapatra} are a class of models that add one more $SU(2)$ gauge group to the  $SU(2)_L \times U(1)_Y$ gauge structure of the SM.  An important prediction of these models is the existence of new heavy $Z'$ and $W'$ gauge bosons, present also in other extensions of the SM \cite{altarelli, PDG}. 
In $G(221)$ models the symmetry  is spontaneously broken twice, giving mass to the
 new $W'$ and $Z'$ bosons and to the SM bosons $W$ and $Z$. 
Depending on the symmetry breaking pattern, there are several versions: left-right symmetric models \cite{naturalLRSM,isoconjugateLRSM,baryon}, lepto-phobic, hadro-phobic, fermio-phobic \cite{phobic1,phobic2,phobic3}, un-unified \cite{uu1} and non-universal \cite{nu1,nu2,nu3}. 
The Higgs sector 
is enlarged with a pair of heavy charged Higgs $H^\pm$, similar to the two-Higgs doublet-models (2HDM) \cite{branco, chen_dawson} or supersymmetric theories \cite{SUSY_H}.  
In some $G(221)$ models, a consequence is the existence of 
a $W'hH^\pm$  interaction, where $h$ stands for a light neutral boson that can be identified with the SM Higgs.

A review of $G(221)$ models and a global analysis of the phenomenological constraints on their parameters from precision data can be found in \cite{hsieh}. 
More recently, the LHC phenomenology of models with an additional $SU(2)$ group   was investigated in  \cite{cao, spethmann, jezo}, while the coupling of $W'$ to a pair of odd heavy Higgs particles was studied in \cite{dobrescu}. 

In the present work,
we study several final states produced in $pp$ interactions at the LHC through the $W' \to hH^\pm$ decay, predicted by a class of $G(221)$ models. The aim was to consider new decay channels of the $W'$ boson, whose searches were based up to now on its decays to $l\bar{\nu_l}$ and quark-antiquark pairs \cite{PDG, atlas, cms, atlasqq}.  As a first estimate, we compared the predictions of $G(221)$ models for several fiducial cross sections with selections adopted in searches for SUSY, with the model-independent upper limits on the visible cross-sections for beyond-standard model physics measured by ATLAS  \cite{0l6jMET,1l6jMET,2l4jMET_sigma,1l4bjMET,susypublic}.

The paper is organized as follows. In section \ref{sec:higgs} we present the Higgs sector of the $G(221)$ models. In section \ref{sec:lagrangian} we briefly discuss the Lagrangian terms of interest for the present study and derive the form of the $W'hH^\pm$ interaction. 
Using recent phenomenological constraints on the $G(221)$ models \cite{hsieh}, the $W' \to hH^+$ decay width  and the cross section for the $hH^+$ inclusive pair production in $pp$ collisions at $\sqrt{s}=$8 and 14 $\tev$ are calculated in section \ref{sec:cross-section}. We also give here the total cross section for the production of a final state  with two leptons, four jets and missing transverse energy in  $pp$ collisions at 8 and 14 TeV.
Section \ref{sec:2l4jMET} is devoted to the study of several final states produced at LHC through the $W'\to hH^+$  decay. The simulation framework and the kinematical cuts applied in the analysis are described. Finally,
 we compare the fiducial cross sections calculated with $G(221)$ models with those predicted by 2HDM and with the model-independent 
upper  limits on the 
visible cross-sections, determined by  ATLAS in SUSY searches based on the same final states.

\section{Higgs sector of $G(221)$ models} \label{sec:higgs}
We adopt the symmetry breaking of the $G(221)$ models  as discussed in \cite{hsieh}
\begin{equation}
SU(2)_1 \times SU(2)_2  \times U(1)_X \rightarrow SU(2)_L \times 
U(1)_Y \rightarrow U(1)_{em}
\end{equation}
by means of two symmetry breaking stages. For the first stage, the hypercharge is $Y = X + T_3^{(2)}$. For the second one, 
$Q=T_3^{(1)} + Y/2=  T_3^{(1)} + (T_3^{(2)} + X)/2 $. This is the scenario of the left-right (LR) symmetric, lepto-phobic, hadro-phobic and fermio-phobic models. Another scenario, proper to the un-unified and non-universal models,  breaks first  $SU(2)_1 \times SU(2)_2$ to $SU(2)_L$, and then the SM reduction  $SU(2)_L \times U(1)_Y \rightarrow U(1)_{em}$ takes place. The Higgs sector for these scenarios may contain a doublet, a triplet, or a bidoublet Higgs.

We shall consider only the first symmetry breaking pattern, because it predicts a nontrivial coupling between a  heavy gauge boson $W'$, a neutral light Higgs boson $h$ and a charged Higgs boson $H^+$.    

For the first symmetry breaking stage we adopt a doublet complex scalar 
\begin{equation}\label{eq:Phi}
\Phi =  \left( \begin{array}{c}
         \phi^+ \\
         \phi^0
        \end{array} \right),
\end{equation}
which transforms as $(1,2,\frac{1}{2})$ under the group action and gives masses to the heavy gauge bosons $W'$ and $Z'$.
For the second symmetry breaking stage, we adopt a bidoublet complex scalar $\mathcal{H}$ transforming as $(2,\bar 2,0)$. For fullfiling the charge conservation relation 
$Q = T_3^{(1)} + (T_3^{(2)} + X $)/2, it must have  the form:
\begin{equation}\label{eq:H}
\mathcal{H} =  \left( \begin{array}{c c}
         h_1^0 & h_1^+ \\
         h_2^- & h_2^0 
        \end{array} \right).
\end{equation}
 The doublet and the bidoublet have nonzero vacuum expectation values (v.e.v.):
\begin{equation}\label{eq:vev1}
 \la \Phi \ra = \frac{1}{\sqrt{2}} \left( \begin{array}{c}
         0 \\
         u
        \end{array} \right),
\end{equation}
and
\begin{equation}\label{eq:vev2}
 \la \mathcal{H} \ra = \left( \begin{array}{c c}
         k & 0 \\
         0 & k'
        \end{array} \right),      
\end{equation}
where $u$, $k $ and $k'$ are real. It is convenient to write the parameters $k $ and $k'$  as \cite{hsieh}
\begin{equation}
\label{eq:betabar}
 k=\frac{v \cos \bar\beta}{\sqrt{2}}, \qquad k'=\frac{v \sin \bar\beta}{\sqrt{2}},
\end{equation} 
where  $v \approx 246\gev$ according to SM.

The symmetry breaking scale of the first stage is much higher than the 
electroweak one, therefore the parameter  \cite{hsieh}
\begin{equation}{\label{eq:x}}
x \equiv \frac{u^2}{v^2}
\end{equation} 
 is expected to be very large.  In the present study we have adoptod from \cite{hsieh} the phenomenological constraint $x\ge 100$. 
As remarked in \cite{hsieh,cao}, the corrections depending on $\bar\beta$ to the observables are numerically suppressed, so this parameter is not constrained by the global-fit analyses. For some of the discussions given below, we assumed that the ratio  $k'/k$ is small. 

The Higgs sector of the model contains 12 real scalars, from which 6 zero-mass modes go into the 
longitudinal degrees of freedom of $W'^\pm$, $Z'$ in the first symmetry 
breaking stage, respectively of $W^\pm$ and $Z$  in the second one. After the 
symmetry breakings,  6 physical degrees of freedom remain, from which we will be interested in a neutral SM-like Higgs  boson $h$ and a charged Higgs boson $H^+$.

\section{ Lagrangian of $G(221)$ models and $W'hH^{\pm}$ interaction} \label{sec:lagrangian}
The Lagrangian of $G(221)$ models is invariant under the transformations $\Phi\to \Phi'$, $\mathcal{H} \to \mathcal{H}'$, where \cite{isoconjugateLRSM}
\begin{equation}\label{eq:U1U2}
 \Phi'=U_2 \Phi, \quad\quad  \mathcal{H}' = U_1 \mathcal{H} U_2^{\dagger},
\end{equation}
with $U_1 \in SU(2)_1$, $U_2 \in SU(2)_2$. We are interested in the kinetic and potential terms 
of the Lagrangian:  
\begin{equation}\label{eq:lagrangian}
\mathcal{L}\sim  \Tr\left[(\mathcal{D}_{\mu} \mathcal{H})^{\dagger}
(\mathcal{D}^{\mu}\mathcal{H}) \right] - V(\mathcal{H},\widetilde{\mathcal{H}},\Phi,\widetilde\Phi) ,
\end{equation} 
where
\begin{equation}
\mathcal{D}_{\mu} \mathcal{H} = 
\partial_\mu \mathcal{H} - i\, \frac{g_1}{2}\, \sum _{j=1}^3\tau_j  W_j \mathcal{H} + i \,
\frac{g_2}{2} \,\mathcal{H^\dagger}  \sum _{j=1}^3\tau_j  W'_j
\end{equation} 
is the covariant derivative that fixes the local gauge interaction, $\tau_j$ are 
the Pauli matrices, and $g_1$, $g_2$ are the coupling constants for the first and the
second symmetry groups. We placed ourselves in the frame of $G(221)$ models which identify the gauge bosons of the groups $SU(2)_1$ and $SU(2)_2$ with the SM boson $W$ and an additional boson $W'$, respectively.

The most general Higgs potential $V$ is written in terms of the fields $ \mathcal{H}$, $\widetilde {\mathcal{H}}$, $\Phi $ and $\widetilde \Phi$, where $\widetilde {\mathcal{H}} = \tau_2  \mathcal{H}^*\tau_2$ and $\widetilde\Phi=i\tau_2 \Phi^*$ ($\mathcal{H}^*$ and $ \Phi^*$ denote the complex conjugates). The important remark here is that  $\widetilde {\mathcal{H}}$ and $\widetilde \Phi$ transform under $SU(2)$   exactly as $\mathcal{H}$ and $\Phi$, respectively  \cite{naturalLRSM, senjanovic, stancu}.  We mention that the original fields in the Lagrangian do not coincide with the physical fields, defined as eigenstates of the mass matrices. Likewise, due to the mixings, the parameters entering $\mathcal{L}$ are not exactly the parameters determined phenomenologically. 
We shall specify the differences in the cases of interest.

\subsection{Higgs potential}
We adopt the potential \cite{senjanovic, kayser} 
\begin{eqnarray}\label{eq:V}
&~&\hspace{-2.9cm} V(\mathcal{H},\widetilde{\mathcal{H}}, \Phi, \widetilde \Phi) = - \mu_1^2 \Tr ( \mathcal{H}^\dagger  \mathcal{H}) +  \lambda_1\, [\Tr  (\mathcal{H}^\dagger  \mathcal{H})]^2 + \lambda_2 \Tr  (\mathcal{H}^\dagger  \mathcal{H}  \mathcal{H}^\dagger 
 \mathcal{H})  \\
&+& \frac{1}{2} \lambda_3 \,[ \Tr ( \mathcal{H}^\dagger \widetilde{\mathcal{H}}) +\Tr (\widetilde{\mathcal{H}}^\dagger \mathcal{H})]^2 + 
\frac{1}{2} \lambda_4 \,[\Tr ( \mathcal{H}^\dagger \widetilde{\mathcal{H}}) -
\Tr (\widetilde{\mathcal{H}}^\dagger  \mathcal{H})]^2 \nonumber \\
&+& \lambda_5 \Tr ( \mathcal{H}^\dagger  \mathcal{H} \widetilde{\mathcal{H}}^\dagger \widetilde{\mathcal{H}})  + \frac{1}{2} 
\lambda_6\, [\Tr (\mathcal{H}^\dagger \widetilde{\mathcal{H}}  \mathcal{H}^\dagger \widetilde{\mathcal{H}}) + \Tr
 (\widetilde{\mathcal{H}}^\dagger  \mathcal{H} \widetilde{\mathcal{H}}^\dagger   \mathcal{H})] \nonumber\\ 
&-& \mu_2^2 \Phi^\dagger \Phi + \rho_1 (\Phi^\dagger \Phi)^2 
+\alpha_1\Tr (\mathcal{H}^\dagger  \mathcal{H}) \Phi^\dagger \Phi + \alpha_2 \Phi^\dagger  \mathcal{H}^\dagger  \mathcal{H} \Phi +
\alpha'_2 \Phi^\dagger \widetilde{\mathcal{H}}^\dagger \widetilde{\mathcal{H}} \Phi,\nonumber
\end{eqnarray}
where $\mu_1$, $\mu_2$, $\lambda_1\,,\ldots \lambda_6$, $\alpha_1$, $\alpha_2$ and $\alpha'_2$ are 
real  parameters. We note that  $\widetilde\Phi$ does not produce new terms in this case. 

The equations for the minimum  of the  potential, which can be written in terms of the v.e.v. of the fields as\cite{senjanovic, kayser}
\begin{equation}\frac{\partial V}{ \partial u}=\frac{\partial V}{ \partial k} = \frac{\partial V}{ \partial k'} = 0, \end{equation}

 do not have an unique solution.  We adopted the following constraints:
\begin{eqnarray}
\mu_2^2 &=& \alpha_1 ( k^2 + k'^2) +\alpha'_2 k^2   + \alpha_2 k'^2 + \rho_1 u^2,  \\
\mu_1^2 &=& 2 (\lambda_1  + \lambda_2)(k^2+ k'^2)  + \frac{1}{2} \alpha_1 u^2 +\frac{ (\alpha'_2 k^2 - \alpha_2 k'^2) u^2 }{2 \Delta k^2 }, \nonumber \\
\lambda_3 &=&  \frac{1}{4} (\lambda_2-\lambda_5 - \lambda_6)  -  \frac{\Delta \alpha u^2 }  {16 \Delta k^2 },\nonumber
\end{eqnarray}
where $\Delta \alpha = \alpha_2 - \alpha'_2$ and $\Delta k^2= k^2 - k'^2$. Using (\ref{eq:betabar}), we can replace $ k^2 + k'^2= v^2/2$. 

By inserting this solution in the expression (\ref{eq:V}) of the potential, with the scalar fields in $\Phi$ and $\mathcal{H}$ replaced by their vacuum expectation values (\ref{eq:vev1}) and (\ref{eq:vev2}), we extract the mass matrix of the charged scalar sector:
\begin{equation}\label{eq:Mpm}
M_+^2 =\left(\begin{array}{ccc}
       \frac{\Delta \alpha k^2 u^2}{2\Delta k^2} &  \frac{ \Delta \alpha k u}{\sqrt{2}} & \frac{\Delta \alpha k k' u^2}{2\Delta k^2} 
         \\
      \frac{ \Delta \alpha k u}{\sqrt{2}} &  \Delta \alpha \Delta k^2   &    \frac{ \Delta \alpha k' u}{\sqrt{2} }        \\
\frac{\Delta \alpha k k' u^2}{2\Delta k^2}     &    \frac{ \Delta \alpha k' u}{\sqrt{2} }    &  \frac{\Delta \alpha k'^2 u^2}{2\Delta k^2}           \\
\end{array}\right),
\end{equation}
in the basis   $h_1^+$,  $\phi^+$, $h_2^+$. The expressions simplify for $k \rightarrow \frac{v}{\sqrt{2}}$, $k' \rightarrow 0$.  In this limit, investigated in many studies  of $G(221)$ models \cite{senjanovic,kayser,fogli1985}, the field  $h_2^+$  is a massless eigenstate absorbed in the charged gauge field. The fields $h_1^+$ from the bidoublet and $\phi^+$ from the doublet are mixed.  After the diagonalization of the mass matrix obtained from (\ref{eq:Mpm}),  we obtain  another  would-be Goldstone boson  and a physical charged Higgs 
boson  $H^+$,  defined as
\begin{equation}\label{eq:Hplus}
H^+ = \frac{1}{\sqrt{u^2 + v^2}}(u h_1^+ + v \phi^+), \quad\quad \quad  m_{H^+}^2  = \frac{1}{2}\Delta \alpha (u^2 + v^2). \nonumber
\end{equation}

For the neutral sector, we consider first the real parts of the fields $h_{1}^0$,  $h_{2}^0$ and $\phi^0$.  In the limit $k\to v/\sqrt{2}$,~ $k'\to 0$, the field  $h_{2,r}^0$ decouples from the other fields.  The remaining $2\times 2$ mass matrix is written in the basis $h_{1,r}^0$, $\phi_r^0$ as
\begin{equation}
M_0^2 =\left(
\begin{array}{ccc}
2(\lambda_1 + \lambda_2) v^2  &   2(\alpha_1 +\alpha_2') u v         \\
2(\alpha_1 +\alpha_2') u v            &             2\rho_1 u^2         \\
\end{array}
\right),
\end{equation}
and   has the eigenvalues
\begin{equation}
m^2_{1,2}=\frac{1}{2}\left(\lambda v^2+\rho_1 u^2 \mp \sqrt{ (\lambda v^2+\rho_1 u^2)^2-4 v^2 u^2 (\lambda\rho_1-\alpha^2)}\right),
\end{equation}
where $\lambda=\lambda_1+\lambda_2$ and $\alpha=\alpha_1 +\alpha_2'$. We define the mass eigenstates $h$ and  $H^0$ through 
\begin{equation}\label{eq:hH0}
h_{1,r}^0 \sim \frac{h}{\sqrt{2}}+\epsilon\, H^0, \quad \quad  \phi_r^0 \sim - \epsilon\, h + \frac{H^0 }{\sqrt{2}},
\end{equation}
where $\epsilon$ is suppressed by the small ratio $v/u$. To leading order in  $v^2/u^2$  the squared masses are
\begin{equation}\label{eq:mhH0}
 m_{h}^2  \sim \frac{v^2}{2}\left(\lambda_1 + \lambda_2 - \frac{\alpha_1 +\alpha_2'}{\rho_1}\right), \quad \quad m_{H^0}^2 \sim 
\frac{\rho_1 u^2}{2}.
\end{equation}
The light  boson $h$, whose mass is proportional to the v.e.v. $v$,  is assumed to be a SM-like Higgs boson.  The neutral Higgs $H^0$ and the charged one $H^+$ are expected to be heavier,  their masses being proportional to the large v.e.v. $u$.  

It is of interest to study also the mass matrix for imaginary fields $h_{1,i}^0$,  $h_{2,i}^0$ and $\phi_i^0$. It turns out that  $\phi_i^0$ is decoupled from the other fields and has zero mass, being absorbed into the gauge bosons degrees of freedom. Moreover,  after the diagonalization of the mass matrix we obtain $h_{1,i}\sim \epsilon'\, A^0+ G^0$ and $h_{2,i}\sim A^0 -\epsilon'\, G^0$, where  $A^0$ is a heavy CP-odd neutral boson, $G^0$ is another Goldstone boson and  $\epsilon'$ is small in the limit   $k'\to 0$. 

A detailed analysis of the Higgs sector of $G(221)$ models, in particular of the $LR$ symmetric ones, was performed in \cite{senjanovic, kayser}, and more recently in  \cite{leejung1, leejung2}. Due to the large number of free parameters, the $G(221)$ models have a large flexibility in the Higgs sector. Therefore, it is possible to adjust the properties of the light boson  $h$ such as to match those of the SM Higgs.  Our purpose here was to show that the physical fields are related in a simple way to the original fields in the Lagrangian, which is important for the derivation of the coupling of interest, between the additional gauge boson $W'$ and a pair consisting of a charged and a neutral Higgs boson. The derivation is presented in the next subsection.

 \subsection{$W'hH^+$  interaction}
The coupling of interest is derived from the kinetic part of the Lagrangian \eqref{eq:lagrangian}, which we write explicitly in terms of the charged gauge fields defined by setting $W^1_\mu = ( W^{+}_\mu + W^-_\mu)/\sqrt{2}$,
 $W^2_\mu = (-W^+_\mu + W^-_\mu)/(\sqrt{2} i)$ and similarly for $W'$. One finds easily a term containing the original fields
$ h_1^+$, $h_{1,r}$ and $W'^-$ of the Lagrangean $\mathcal{L}$. To pass to the physical fields we  use the relations  $h_1^+ \sim H^+$ and $h_{1,r}^0\sim h/\sqrt{2}$ which follow from (\ref{eq:Hplus}) and (\ref{eq:hH0}) to first order in $v/u$. We also use the $W'$-$W$ mixing term  \cite{hsieh}
 \begin{equation}\label{eq:WW'}
  \mathcal{L} \sim -\frac{1}{2}\, g_1 g_2 v^2 \cos \bar\beta \, \sin \bar\beta \,W^-_\mu 
W'^{+\mu} +h.c.,
 \end{equation} 
 which is obtained from  \eqref{eq:lagrangian} by  replacing  $h_1^0$ and  $h_2^0$ by their v.e.v. from (\ref{eq:vev2}) and 
(\ref{eq:betabar}) and collecting the terms proportional to  $W^-_\mu W'^{+\mu}$.
For small $\sin \bar\beta$, {\em i.e.} for $k'\to 0$, the  $W$-$W'$ mixing given in (\ref{eq:WW'}) is small, therefore the physical field $W'$, defined as in Ref. \cite{hsieh} from the diagonalization of the $W$-$W'$ mass matrix, coincides practically with  the original  $W'$ field in the Lagrangian.
Thus,  we obtain the $W'^\mp hH^\pm$ interaction written in terms of the physical fields as
\begin{equation}\label{eq:W'}
\mathcal{L} \sim
 -\frac{1}{2} \,i g_2  W'^-_\mu ( h\, \partial^{\mu}H^+ - H^+\,\partial^{\mu} h)  + h.c.
 \end{equation}

It is of interest to investigate also other interactions of $W'$ in the adopted model. One possible coupling is that of $W'$ to the pair $H^0 H^+$, where $H^0$ is a heavy neutral Higgs boson. It arises from the same  term  $W'h_{1,r} h_1^+$  of the Lagrangean $\mathcal{L}$, which yielded the coupling $h W' H^+$ of interest. However, from  (\ref{eq:hH0}) it follows that the contribution of $H^0$ to the field   $h_{1,r}$ is suppressed by the small ratio $v/u$. Moreover, it turns out that the coupling of  $W'$ to the field $h_{2,r}^0$, which generates another Higgs field, is identically 0. 

In the imaginary fields sector, one can check that the only nonzero term in $\mathcal{L}$  is $W'h_{1,i} h_1^+$, while  the coupling $W' h_{2,i} h_1^+$ vanishes identically. Using the expression of $h_{1,i}$ in terms of physical fields given the previous subsection, we obtain from the term $W'h_{1,i} h_1^+$ a physical coupling $W'A^0H^+$  suppressed in the limit $k'\to 0$\footnote{We note that in the model presented in \cite{dobrescu} the opposite situation occurs, {\em i.e.}  $W'$ interacts with a CP-odd boson and  a heavy CP-even boson, and has a suppressed coupling with the SM-like Higgs boson $h$.}.

The couplings of $W'$ to the pair $Wh$ or to the SM bosons $WZ$ can proceed  through the   $W'W$ mixing. As shown in \cite{cao}, for models with the first breaking symmetry pattern adopted here, these couplings are suppressed
  by the large parameter $u/v$  and vanish identically for  $k'=0$. The couplings of $W'$  to fermions depend on the details of the 
model. As discussed in \cite{cao},  the Sequential Standard Model (SSM), where 
$W'$ has the
same couplings to fermions as the SM $W$ boson, can be considered as a reference for the $G(221)$ models, since the gauge boson production cross sections are obtained from the corresponding quantities in the SSM  by a simple scaling with a factor depending on the couplings.

 \subsection{Yukawa  interactions}
Due to the complexity of the Higgs sector, predictions on the interactions between fermions and the additional Higgs bosons are not available in the general frame of $G(221)$ models. As remarked in \cite{hsieh}, in these models there are many free parameters in the Yukawa
sector, which can lead to interesting flavor phenomena, particularly in the
arena of neutrino physics  (see, for example \cite{mohapatra}).

 More detailed predictions are possible in specific models like the left-right symmetric models, where the groups $SU(2)_1$ and 
$SU(2)_2$ are identified with $SU(2)_L$ and $SU(2)_R$, respectively. According to \cite{leejung1, leejung2}, the structure of the Yukawa couplings of the bidoublet Higgs field ${\mathcal H}$ in such models is quite different from that of the Higgs fields of 2HDM. For instance, according to Eq. (15) of \cite{leejung2}, the ratio of the $H^+ \bar t b$ squared couplings in LR-symmetric models and type II 2HDM \cite{branco, chen_dawson}  writes as:
\begin{equation} \label{eq:ratio}
\frac{g^2_{\rm LR}}{g^2_{\rm 2HDM}}\approx \frac{m_t^2[ (1 + \xi^2)^2 /(1 - \xi^2 )^2 + 4\xi^2 /(1 - \xi^2)^2]}{m_b^2 \tan^2\beta + m_t^2 \cot^2\beta},
\end{equation}
where  $\xi\equiv k'/k$ in our notation (\ref{eq:vev1}),  and $\beta$ is the mixing parameter of 2HDM. The ratio (\ref{eq:ratio}) is larger than unity
 except for very small and very large values of  $\tan\beta$, and the result is stable for small $\xi \le 0.1$. 

\section{Decay width $\Gamma(W'\to hH^+)$}\label{sec:cross-section}
For simplicity, in what follows we shall refer to the $W'$ decay into the positively charged Higgs. Using \eqref{eq:W'}, we write the amplitude of this process at tree level as
\begin{equation}\label{eq:M}
 \mathcal{M} = \frac{i g_2}{2} \epsilon'_{\mu}(p_1-p_2)^{\mu},
\end{equation}
where $\epsilon'$ is the $W'$ polarization 4-vector and $p_1$, $p_2$ are the  momenta of the two final
Higgs bosons.  We note that angular momentum conservation implies that the two final bosons are produced in a state of orbital momentum 1 in the $W'$ rest frame.

The differential decay rate  is given by 
\begin{equation}\label{eq:dGamma}
 \rmd \Gamma(W'\to hH^+) = |\mathcal{M}_{av}|^2  \frac{1}{2 m_{W'}} 2 \pi^4 
\delta^4 (P - p_1 - p_2 ) \frac{d^3 p_1 d^3 p_2}{(2 \pi)^3  2E_1  (2 
\pi )^3  2E_2}\,, 
\end{equation}
where the squared  amplitude averaged over the initial $W'$ polarization states is
\begin{equation}\label{eq:aver}
 |\mathcal{M}|_{av}^2 =\frac{g_2^2}{4}\, \frac{1}{3} \sum_{\epsilon'} \epsilon'_{\mu} 
\epsilon'^{\nu} (p_1 - p_2 )^{\mu} (p_1 - p_2)_{\nu}. 
\end{equation}
Using
\begin{equation}
 \sum_{\epsilon'} \epsilon'_\mu \epsilon'^\nu = -g_{\mu \nu} + \frac{P_\mu 
P_\nu}{m_{W'}^2}, \quad P=p_1+p_2, 
\end{equation}
to evaluate (\ref{eq:aver}) and performing the trivial phase space integral of (\ref{eq:dGamma}), we finally obtain the partial width as
\begin{equation}\label{eq:gamma}
\Gamma (W'\to hH^+) =  \frac{g_2^2} {192 \pi}\, \frac{\lambda^{3/2}(m^2_{W'}, m^2_{H^{+}}, m^2_{h})}{m_{W'}^5}, 
\end{equation}
in terms of the standard kinematical function
\begin{equation}
\lambda(a,b,c) = a^2+b^2+c^2-2 ab-2ac-2bc.
\end{equation}

A constraint on the coupling $g_2$ entering (\ref{eq:gamma}) can be derived by writing the  mass of $W'$ as \cite{hsieh}
\begin{equation}\label{eq:massW'}
 m^2_{W'} =   \frac{1}{4}  g_2^2 u^2 + \frac{1}{4}  g_2^2 v^2   + \frac{ g^2_1 v^4 \sin^2 2\bar\beta}{4 u^2},
\end{equation} 
 where the first two terms are the contributions to the mass  after the first and second symmetry breaking 
stages, and the third term is a correction from the $W'$-$W$ mixing (\ref{eq:WW'}). By  neglecting the last term due to the large $u$ in the denominator  and using the ratio defined in (\ref{eq:x}),  we write \eqref{eq:massW'} as
\begin{equation}
 m_{W'}^2 = \frac{1}{4}g_2^2  v^2 (x + 1),
\end{equation}
from which we obtain, to leading order in the large parameter $x$,
\begin{equation}\label{eq:g2} g_2 \approx \frac{2m_{W'}}{v \sqrt{x}}.\end{equation}     
Using this estimate in \eqref{eq:gamma}, we obtain 
\begin{equation}\label{eq:width}
 \Gamma(W'\to hH^+) = \frac{\lambda^{3/2}(m^2_{W'}, m^2_{H^+}, m^2_{h}) }{48 \pi x v^2  m^3_{W'}}= \frac{G_F}{ 24\sqrt{2} \pi x} \frac{\lambda^{3/2}(m^2_{W'}, m^2_{H^+}, m^2_{h}) }{ m^3_{W'}}.
\end{equation} 
 
\begin{table}[htb]
\caption{\label{tab1} Partial width  $\Gamma(W' \to hH^+)$, in GeV, calculated from (\ref{eq:width}) as a function of $m_{W'}$, for three value of the parameter $x$ and $m_{H^+}=300\,(500) \gev$.  }
\vspace{0.3cm} 
\begin{center}
\begin{tabular}{|c| r r| r r |r r |} \hline
\multirow{2}{*}{$m_{W'}$[GeV]}    & \multicolumn{2}{c|}{$x=100$} &      \multicolumn{2}{c|}{$x=500$}      &     
\multicolumn{2}{c|}{$x=1000$} \\\cline{2-7}
   &  \multicolumn{1}{c}{~~300}   &   \multicolumn{1}{c|}{~~500}    & 
\multicolumn{1}{c}{~~300}     &   \multicolumn{1}{c|}{~~500}  & \multicolumn{1}{c}{~~300}  &   \multicolumn{1}{c|}{~~500}         \\ \hline
    1000 &  0.77 & 0.41 & 0.15 &0.08 & 0.08   &0.04 \\
    1500 &  3.16 &2.49 &0.64 & 0.50& 0.32& 0.25  \\
    2000 & 8.00 & 7.05 &1.61  &1.42 & 0.81   & 0.71  \\ 
    2500 & 16.10 & 14.87 &3.25 & 2.99 &1.62   & 1.50   \\
    3000 & 28.27 & 26.77  &5.69 &5.39 &2.85 &2.70  \\
    3500 & 45.32 & 43.55  &9.14 &8.78 &4.57 & 4.39    \\
    4000 & 68.07 & 66.03 &13.72 & 13.31 &6.87  & 6.66  \\ 
    4500 & 97.32  &  95.02 &19.63 &19.16 &9.82 &9.59   \\
    5000 & 133.90 & 131.33 &26.99 &26.48 &13.51 &13.25    \\ 
     \hline
\end{tabular}\end{center} 
\end{table}

\begin{figure}[htb]\begin{center}\vspace{0.8cm}
\includegraphics[width=10cm]{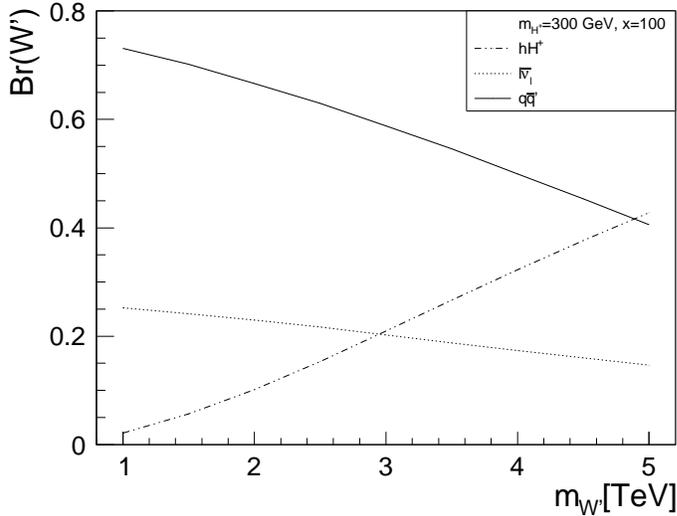}
\caption{$W'$ branching fractions, including the $W'\to hH^+$ decay.  }
\label{fig:br}\end{center}
\end{figure}
 
 Several values of $\Gamma(W'\to h H^+)$ calculated using (\ref{eq:width}) are given in  Table \ref{tab1}. We adopted  the lower 
bound $ x\ge 100$ derived from  recent  phenomenological
studies \cite{hsieh} of $G(221)$ models. The mass of the light Higgs was set to $m_{h}=126\gev$.
 Lower limits on the  $W'$ mass  were obtained recently at LHC from the investigation of $W'$ decays into leptons  \cite{atlas, cms} 
and quark-antiquark pairs \cite{atlasqq}.
The most 
stringent limit, of 3.8 TeV, is set by the CMS Collaboration  \cite{cms}, by assuming  a “sequential” $W'$, which has the same couplings to quarks and leptons as the SM $W$ boson.  However, when  other models and other possible decays of $W'$ are considered, the existing limits may be relaxed \cite{dobrescu}.
Therefore, in our study we considered a larger interval of masses, between 1 and 5 TeV. As for the mass of the charged Higgs boson, limits have been derived recently at LHC \cite{chargedHiggs}. Also, stringent constraints
on  $m_{H^+}$  were derived from $B$ decays in the frame of type II 2HDM 
by Belle and BABAR \cite{horii}. We adopted for $m_{H^+}$ two values compatible with these limits. 

\begin{figure}[htb]\begin{center}\vspace{0.8cm}
\includegraphics[width=10cm]{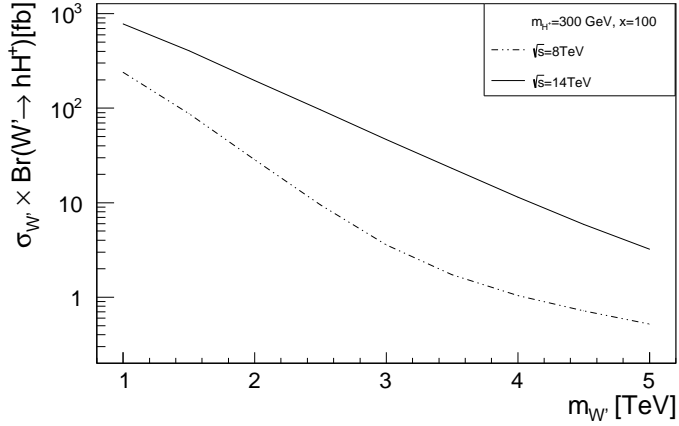}
\caption{ Cross section (in fb)  of inclusive production of a  $hH^+$ pair via the $W' \to hH^+$ decay in $pp$ collisions at  8 and 14 TeV.  }
\label{fig:s1}\end{center}
\end{figure}

In Fig. \ref{fig:br} we present the branching fractions ${\rm Br}(W' \to XY)$, using for the leptonic and quark channels the SSM \cite{altarelli}. The $q\bar{q'}$ channel includes $t\bar b$. As discussed in the previous section, other decays of $W'$ possible in the frame of $G(221)$ models, like $W'\to H^0 H^+$, $W'\to A^0 H^+$, $W'\to WZ$ or $W'\to hW$ are suppressed and can be neglected in the approximation of large $u/v$ and  small $k'$. 

We computed also the inclusive cross section for $hH^+$ pair production in $pp$ collisions,  given by $\sigma_{prod}^{W'} \times {\rm Br}(W'\to hH^+) $.  The $W'$ production cross section in $pp$ collisions was calculated in \cite{sullivan} for various couplings at NLO in QCD. We computed $\sigma_{prod}^{W'}$ with  PYTHIA LO SSM implementation, with the parton distribution functions for the proton set to CTEQ 5L \cite{PYTHIA6.4.4}. The results are presented in Fig. \ref{fig:s1}, where we show the  cross section for the $h H^+$ production in $pp$ collisions at $\sqrt{s}=$ 8 and 14 TeV,  for $m_{H^+}=300 \gev$ and $x=100$. 

\begin{figure}[htb]\begin{center}
\includegraphics[width=7cm]{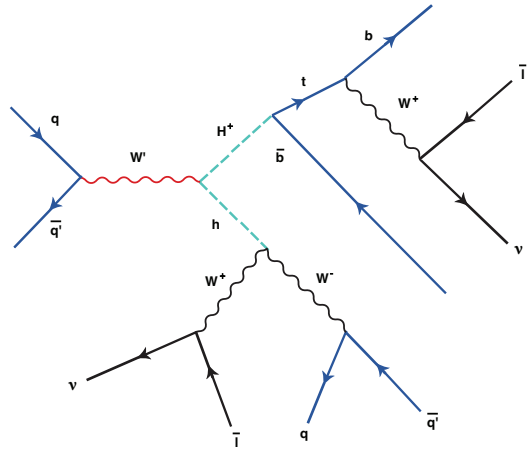}
\caption{ Production of the  final state with 2 same-sign leptons, 4 jets and ${\rm E_T^{miss}}$   in $pp$ collisions via the $W' \to hH^+$ decay.  }
\label{fig:Feynman}\end{center}
\end{figure}

\begin{figure}[htb]\begin{center}\vspace{0.8cm}
\includegraphics[width=10cm]{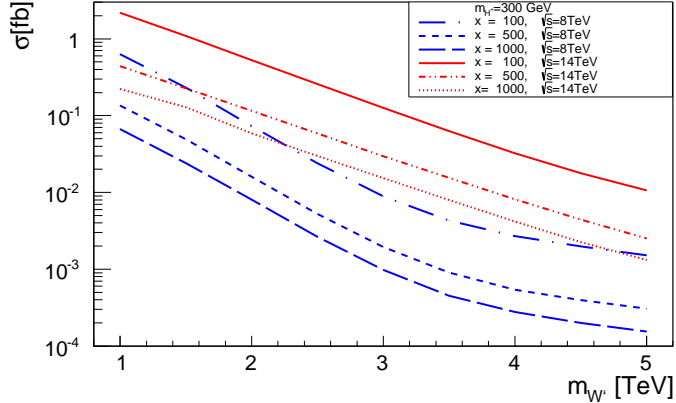}
\caption{Cross section $\sigma$ (in fb) for the  production of 2 electrons, 4 jets and $E_{\rm T}^{\rm miss}$ via the $W' \rightarrow hH^+$ channel in $pp$ collisions at 8 and 14 TeV. }
\label{fig:sfinal}\end{center}
\end{figure}

For illustration, we consider also the production of 2 leptons, 4 jets and ${\rm E_T^{miss}}$ in $pp$ collisions via the $W' \to hH^\pm$ decay (one possible diagram  is shown in Fig. \ref{fig:Feynman}). In Fig. \ref{fig:sfinal} we present the total cross section for the production of this  final state, calculated with the  PYTHIA 6.4 in which we implemented the new channel $W'\to hH^\pm$. Details on the simulations will be given in the next section.

\section{Final state comparison study}\label{sec:2l4jMET}
For several final states, we present a comparison study between the predictions of $G(221)$ models and the experimental upper limits on the visible cross sections determined by ATLAS in SUSY searches. We first describe the simulation procedure and then discuss the specific final state analysis.
\subsection{Simulation framework}\label{sec:sim}
The simulation and the analysis were based on a framework \cite{infrastructura} that chains different software packages together:  Monte Carlo generators, programs for the simulation of the events through the detector and programs for data analysis.
Using PYTHIA 6.4 as Monte Carlo event generator, we produced data samples for final states which comprise leptons, jets and missing transverse energy. The parton distribution  functions set for the proton was CTEQ 5L and the center of mass energy for the $pp$ collisions was set to $8\tev$. 

 As we mentioned above,  PYTHIA 6.4 adopts as default the SSM,  where the $W'$ boson has the 
same couplings to fermions as the  SM boson $W$. We have implemented in PYTHIA 6.4 the additional $W'\to hH^+$ channel, with 
the decay width  and the branching ratio set according to the calculations presented in Section 4. 
  We  allowed $h$ to decay to $WW$, $ZZ$ and $b \bar{b}$, while the charged Higgs  was set to decay as $H^+ \to t \bar b$ in all cases. The top mass was fixed at $172.5$ GeV and the  light Higgs mass $m_{h}=126 \gev$. All the calculations in this section were performed with  $m_{W'}=1 \tev$, $m_{H^+}=300 \gev$ and $x=100$.

 The default coupling for $H^+ \to t \bar b$ decay in PYTHIA 6.4 is the 2HDM  coupling with $\tan \beta=5$,  which may  
underestimate by a large factor the $G(221)$  couplings, as discussed in \cite{leejung1,leejung2} in the context of the left-right symmetric models. In order to work with a more realistic coupling, we performed the simulations in PYTHIA with the choice $\tan \beta=1$, when the ratio
given in (\ref{eq:ratio}) is close to 1 for small values of $\xi$. For other decays involved in the production of the final states we assumed SM couplings.    

For the comparison with the published results of the ATLAS Collaboration, 
we  used the Delphes framework \cite{delphes} which provides a realistic 
fast simulation of the ATLAS detector \cite{ATLAS} and delivers reconstructed physics objects, such as leptons, jets, photons and  missing energy. The data analysis was performed with ROOT \cite{root}.

\subsection{Fiducial cross section comparison for various final state topologies}

The aim of this section is to compare the predictions of the $G(221)$ models with the recently published results from  SUSY searches by the ATLAS Collaboration \cite{0l6jMET,1l6jMET,2l4jMET_sigma,1l4bjMET}.
Several topologies were investigated in these searches: events containing no leptons, six jets and missing transverse energy $E_\text{T}^{\text{miss}}$ \cite{0l6jMET}, one lepton, six jets and $E_\text{T}^{\text{miss}}$ \cite{1l6jMET}, two opposite-sign leptons, four jets and $E_\text{T}^{\text{miss}}$ \cite{2l4jMET_sigma} and one lepton, four b-jets and $E_\text{T}^{\text{miss}} $\cite{1l4bjMET}. 

In the frame of $G(221)$ models, the same states can be produced in $pp$ collisions via the intermediate $W'\to hH^+$ decay, with $h \to W^+ W^-$, $ZZ$ or $b\bar{b}$,  and  $H^+ \rightarrow t \bar{b}$. The total cross sections  for  the final states with 0l+ 
6j+$E_\text{T}^{\text{miss}}$, 1e+6j+$E_\text{T}^{\text{miss}}$, 2e+4j+$E_\text{T}^{\text{miss}}$ and 
1e+4bj+$E_\text{T}^{\text{miss}}$, obtained through the production and decay of $W'$ with the parameters specified above,  are 0.92, 3.96, 0.63 and 10.94 fb,  respectively.

 Of interest are the cross sections obtained after suitable kinematical cuts, which suppress the background and favor the signal.
In the ATLAS studies \cite{0l6jMET, 1l6jMET, 2l4jMET_sigma, 1l4bjMET}, the kinematical cuts imposed on the final state particles were designed such as to favor SUSY searches. These cuts may be non-optimal for $W'$ detection, where all the final states originate from a high-energy particle of large mass. In fact, it turns out that a small number or even no events generated by the simulations based on $G(221)$ models remain after applying the SUSY inspired selections. 

In our study, the kinematical cuts on $p_\text{T}$, $E_\text{T}^{\text{miss}}$ and the pseudorapidity $\eta$ are only a part of the ATLAS SUSY searches set of conditions. 
No cuts were applied on the transverse mass $m_{\text T}$, on the inclusive effective mass $m_{eff}^{inc}$, which is the sum of the transverse momenta of the jets and leptons and the $E_\text{T}^{\text{miss}}$, or on the ratio between transverse mass and effective mass ($m_\text{T}$/$m_{eff}$). There are also several other variables, like the cotransverse mass $m_\text{CT}$ of two $b$-jets, the $p_{\text T}$ scalar sum $H_{\text T}$ of all jets, the invariant mass of  two leptons coming from a $Z$ boson, or the angle between two leptons, which were not constrained in our analysis.  Of course, more kinematical cuts applied on the phase space and the consideration of efficiency reconstruction will further reduce the fiducial cross sections.

 The zero leptons, six jets and $E_\text{T}^{\text{miss}}$ SUSY final state cross-section is set for tight conditions 
\cite{0l6jMET}. Thus we require at least six jets with $p_{\rm T}$ greater than 130 GeV for the first one, and greater than 60 GeV for the other five, and $ E{\rm _T^{miss}}$ greater than 160 GeV. The main source of this signature is the pair production of light squarks, each of whom decays through an intermediate chargino to a quark, a $W$ boson and the lightest neutralino, in the MSUGRA/CMSSM model. 
 
The one lepton, six jets and $E_\text{T}^{\text{miss}}$ channel is considered for binned hard-single lepton channel \cite{1l6jMET}. The number of leptons is exactly one, whereas the number of jets at least six. We set $E{\rm _T^{miss}}>350\gev$, lepton $p_\text{T}$ greater than 25 GeV, and the $p_{\rm T}$ of jets greater than 80 and 50 GeV for the first two, respectively greater than 40 GeV for the other jets. This state appears in the gluino  inspired MSUGRA  model, where a pair of gluinos decay to quarks, a $W$ boson and a neutralino.   

For the final state with two opposite-sign leptons, four jets and $E_\text{T}^{\text{miss}}$ we require at least two isolated opposite-sign leptons and four jets, with $p_\text{T}$ of the leading leptons greater than $25$ GeV and $p_\text{T}$ of the leading jets greater than $30$ GeV \cite{2l4jMET_sigma}. 
This final state was studied in the Gauge Mediated Symmetry Breaking (GMSB), where stop quark is decaying to top and neutralino.
Because the neutralino is not considered the lightest SUSY particle (LSP), it could decay into a $Z$ or the  SM-like Higgs boson and a gravitino.  

The last final state considered consists of one lepton, four $b$-jets and $ E{\rm _T^{miss}}$ \cite{1l4bjMET}. We require exactly one lepton with $p_{\rm T}$ greater than 25 GeV, and at least four jets with $p_{\rm T}$ greater than 80 GeV.  The $ E{\rm _T^{miss}}$ has to be greater than 250 GeV. This state is the feature of gluino to quark-antiquark and neutralino decay models.  
The pseudorapidity regions for the leptons and jets are always $|\eta_{leptons}| < 2.47$ and $|\eta_{jets}|<2.7$, according to the ATLAS detector acceptance. 

In Table \ref{tab3} we present the fiducial $G(221)$ model cross-sections, $\sigma_{fid}= \sigma_{prod} \times A $, where $\sigma_{prod}$ is the production cross-section and $A$ is the acceptance of the detector, which includes the kinematical cuts over the phase space.
 The comparison with the total cross sections given above shows the drastic effect of the kinematical cuts on the 
final states produced through the decay of $W'$.
For completeness, the cross sections corresponding to the channel $pp \to W \to hH^+$ calculated with PYTHIA 6.4 in the frame of 2HDM are also shown.  We indicated separately the contributions of the channels with $WWW$, $ZZW$ and  $4bjW$ intermediate states,   obtained from $h$ decaying to $WW$, $ZZ$ and $b \bar{b}$, respectively. The fiducial cross sections predicted by the $G(221)$ models are considerably larger than those predicted by  2HDM with the default PYTHIA 6.4 parameters.

In  Table \ref{tab3} we give also  the observed  95\% CL upper limits on the visible cross section  $\sigma_{\rm vis}^{\rm obs} $ of beyond SM processes, determined by ATLAS from experimental measurements with selections for SUSY searches.
These are much larger than the fiducial cross sections of the processes involving the production and decay of $W'$.  However, this result  may be due to the use of selections that are not optimal for $W'$ detection. Moreover, the observed limits on the  visible cross sections offer only a model independent  indication on the magnitude of new physics contributions. Detailed studies are necesary to identify proper
kinematical cuts for the final states produced by the  $W'$ decay and to set limits on model parameters from measured data and the SM background in suitable signal regions. 

\begin{table}[thb]
\caption{\label{tab3}Fiducial cross sections (in fb)  with selections from SUSY searches, calculated in $G(221)$ for
$m_{W'}=1000\gev,\, m_{H^+}=300\gev, \, m_h=126\gev$ and $x=100$. For comparison, the 2HDM predictions for the same final states are also shown. In the second column  we give the observed 95\% CL upper limits on new physics cross sections derived by ATLAS in SUSY searches \cite{0l6jMET,1l6jMET,2l4jMET_sigma,1l4bjMET}. }
\scriptsize
\begin{tabular}{cccccccc} \br
Final state & ATLAS   & $G(221)_{WWW}$  & $G(221)_{ZZW} $  
&  $G(221)_{4bjW}$ &  2HDM$_{WWW} $ &  2HDM$_{ZZW} $ &  
2HDM$_{4bjW}$  \\ \mr
$0l,\,6j,\, E_\text{T}^{\text{miss}}  $ & $0.41$ &- & 
$3.11\cdot 10^{-3}$ & - &   - & $6.05 \cdot 10^{-6}$ & -
 \\ \mr
$e, 6 j, \, E_\text{T}^{\text{miss}} $ &$0.33$ & $3.01 \cdot 
10^{-4}$ & $1.08 \cdot 10^{-5}$ & - &    $3.08 \cdot 
10^{-7}$ & $3.8 \cdot 10^{-8}$ & -\\
$\mu, 6 j,   \, E_\text{T}^{\text{miss}}   $ &  $0.35$ & $1.67 
\cdot 10^{-4}$ & $2.05 \cdot 10^{-6}$ & - &  $3.07 \cdot 
10^{-7}$ & $8\cdot 10^{-9}$ &- \\ \br  
$ee,4j,  \, E_\text{T}^{\text{miss}}  $ & $0.17$ & $2.57 \cdot 
10^{-2}$ & $1.29 \cdot 10^{-2}$ & -  & $2.07 \cdot 10^{-4}$ 
& $5.94 \cdot 10^{-5}$ & - \\
$e \mu, 4 j,   \, E_\text{T}^{\text{miss}}  $ & -  & - & - &  - &  - & - &- \\ 
$\mu\mu, 4j,  \, E_\text{T}^{\text{miss}}   $  & $0.17$ & $1.08 
\cdot 10^{-2}$ & $5.49 \cdot 10^{-3}$ & - &   $8.47 \cdot 
10^{-5}$ & $2.85  \cdot 10^{-5}$  & - \\ \br
$e,4bj,  \, E_\text{T}^{\text{miss}}  $  &  $1.69$ & -&- 
&$7.57 \cdot 10^{-3}$ &   - &- & $1.46 \cdot 10^{-5}$  
\\
$\mu, 4bj, \, E_\text{T}^{\text{miss}} $  & $1.09$  & - &- 
&$5.56 \cdot 10^{-3}$ &  - &- & $2.14 \cdot 10^{-5}$   
\\ \br
\end{tabular}
\end{table}

\section{Summary and conclusions}\label{sec:summary}

In this paper we studied the $W'\to hH^+$ decay predicted by some $G(221)$ models \cite{naturalLRSM, isoconjugateLRSM,prD23, baryon, phobic1, phobic2, phobic3, senjanovic, nu2, fogli1985, kayser, uu1, fogli1991, nu1, nu3, mohapatra, leejung1,leejung2, hsieh, cao}.  The aim was to compare the predictions of $G(221)$ models for several fiducial cross sections  with the model-independent upper limits derived by ATLAS in SUSY searches based on the same final states.

We considered $G(221)$ models with two-stage symmetry breaking, with a scalar sector consisting of a complex doublet in the first stage and of a complex bidoublet in the second. 
Due to the large number of free parameters, the $G(221)$ models have a great flexibility in the Higgs and Yukawa sectors. Therefore, the properties of the light neutral Higgs boson $h$ can be adjusted to match the properties of the SM Higgs. 
In the present study we adopted the phenomenological constraint $x\ge 100$ for the parameter defined in (\ref{eq:x})  and the assumption that the ratio   $k'/k$ of the vacuum expectation values  (\ref{eq:vev2}) is small.  In this general frame, we calculated the coupling between a heavy charged gauge boson $W'$, the light neutral SM-like Higgs boson $h$  and a charged non-standard Higgs boson $H^\pm$. We also calculated the partial width $\Gamma(W'\to hH^+)$, the $W'$ branching fractions including the $W'\to hH^+$ channel and the cross section for the inclusive production of the $hH^+$ state in $pp$ collisions at 8 and 14 TeV.

We considered also several specific final states  produced in $pp$ collisions at the LHC through the $W'\to hH^+$ decay.
We  used a simulation framework \cite{infrastructura} that chains PYTHIA 6.4 Monte Carlo generator, the Delphes framework for a fast simulation of the ATLAS detector and ROOT for the data analysis. 
The branching ratios and the total cross-sections for the final states were obtained at LO with PYTHIA 6.4, where we have implemented the new decay channel $W'\to hH^+$ predicted by $G(221)$ models.  
The analysis involved specific $p_\text{T}$, $\eta$ and $E_\text{T}^{\text{miss}}$ selection cuts that were employed by the searches for supersymmetry performed by the ATLAS Collaboration \cite{0l6jMET,1l6jMET,2l4jMET_sigma,1l4bjMET,susypublic}. 
Our study shows that, assuming specific kinematical cuts that were optimized for SUSY searches by the ATLAS Collaboration, $G(221)$ model fiducial cross sections are larger than those predicted by 2HDM, but  considerably below the ATLAS model independent upper limits on new physics cross sections in the corresponding signal regions.  Further studies are necessary in order to identify 
proper kinematical cuts for the final states produced by the decay of $W'$ and to set limits on model parameters from measured data and the SM background in suitably defined signal regions.

\ack{We would like to thank Bogdan Dobrescu for pointing us to the $G(221)$ models and Julien Maurer for very useful discussions and suggestions. This work was supported by the Sectorial Operational Programme Human Resources Development (SOP HRD), financed from the European Social Fund and by the Romanian Government under the contract number SOP HRD/107/1.5/S/82514, and by the Romanian Ministry of Education through the ATLAS - Capacities/Module III CERN project.}

\end{document}